\documentclass[aps,prb,reprint,superscriptaddress]{revtex4-2}
\usepackage{lipsum,amsmath,graphicx,comment,xcolor}

\begin{document}


\title{Thermally-induced mimicry of quantum cluster excitations and implications for the magnetic transition in FePSe$_{3}$}

\author{H. Lane}
\email[]{harry.lane@manchester.ac.uk}
\affiliation{School of Physics and Astronomy, University of St Andrews, North Haugh, St Andrews, Fife KY16 9SS, United Kingdom}
\affiliation{Department of Physics and Astronomy, University of Manchester, Oxford Road, Manchester M13 9PL, United Kingdom}
\affiliation{University of Manchester at Harwell, Harwell Campus, Didcot, Oxfordshire OX11 0FA, United Kingdom}
\author{M. Mourigal}
\affiliation{School of Physics, Georgia Institute of Technology, Atlanta, Georgia 30332, United States}


\date{\today}

\begin{abstract}
In two dimensional magnets, the interplay of thermal fluctuations and spin anisotropy control the existence of long-range magnetic order. In the van der Waals antiferromagnets FePX$_3$, orbital degeneracy in the $t_{2g}$ levels of the Fe$^{2+}$ ions in octahedral coordination yields strong uniaxial anisotropy, which stabilizes magnetic order up to $T \approx 100$ K. Recent inelastic neutron scattering measurements around the magnetic ordering transition have shown the existence of a broad spectrum of magnetic fluctuations with nontrivial momentum dependence, what has been interpreted as evidence for localized entangled cluster excitations. In this paper, we offer an alternative interpretation using classical non-linear spin dynamics simulations. We present stochastic Landau Lifshitz dynamics simulations that reproduce the neutron scattering measurements of \textcite{Chen24:9} on  FePSe$_3$. These calculations faithfully explain the dynamical structure factor's momentum and energy dependence and point to a classical origin for the excitations observed in neutron spectroscopy and that the order-disorder transition can be understood in terms of thermal fluctuations overcoming the anisotropy energy.
\end{abstract}
\maketitle

\section{Introduction}

The electronic and magnetic phase transitions of quantum materials are important proving grounds for quantum and statistical physics. The critical properties of magnetic insulators are often governed entirely by the dimensionality of the lattice and spin spaces of their underlying exchange Hamiltonian~\cite{Bramwell93:73}. Indeed, for systems with isotropic spin-space (Heisenberg spins) in fewer than three spatial dimensions, thermal fluctuations destroy long-range magnetic order~\cite{Mermin66:17}. The magnetic ordering temperature, if it exists, is thus typically dictated by the strength of anisotropy terms and/or inter-layer (or inter-chain) exchange coupling terms~\cite{Kim19:116,Kim20:20,Torelli18:6}. The stabilization of high-temperature magnetic ordering in truly two-dimensional systems is a highly coveted prize in the design of novel heterostructures, devices, and ultra-thin magnetic storage media, and hence, the study of two-dimensional magnets with spin-orbit coupling and crystallographic anisotropy has become a vibrant area of experimental and theoretical investigation~\cite{Zhao24:14,Burch18:563,Wang22:16,Gibertini19:14,Hejazi20:117}.

In (quasi-)two-dimensional magnets with strong spin uniaxial anisotropy, the Ising-like behavior is well-described by Onsager's exact solution~\cite{Onsager44:65}. Such systems fall outside the scope of the Mermin-Wagner theorem due to their discrete spin-orientation symmetry and exhibit a magnetic phase transition. Typically, material realizations lie intermediate between the Ising and Heisenberg limits. Thus, intense theoretical and experimental activities are devoted to understanding the behavior of two-dimensional magnets in this intermediate regime, with examples ranging from floating phases~\cite{Chandra07:40,Barber88:51} and binding of vortex-antivortex pairs ~\cite{Kosterlitz73:6,Kosterlitz17:89} to topological magnons~\cite{McClarty22:13,Pershoguba18:8,Owerre16:28}. The dynamical behavior of spin-anisotropic two-dimensional magnets at temperatures greater than the anisotropy energy but smaller than the exchange energy is intriguing, because short-range spatial correlations -- enhanced by the large coordination number of two-dimensional lattices -- coexist with thermally activated fluctuations.

The MPX$_{3}$ family of antiferromagnetic compounds offers a flexible platform to investigate the role of spin-anisotropy on (quasi-)two-dimensional magnetism~\cite{Grasso02:25,Chittari16:94}, since the divalent metal M$^{2+}$ that sits at the vertices of the honeycomb lattice can be occupied by Co$^{2+}$~\cite{Kim20:102,Wildes17:29,Wildes23:107}, Ni$^{2+}$~\cite{Kim19:10,Wildes15:92,Wildes22:106,Scheie23:108}, Mn$^{2+}$~\cite{Okuda86:55} or Fe$^{2+}$~\cite{Coak21:11,Wildes20:127,Lancon16:94}. This opens the possibility of controlling spin-space dimensionality (e.g. integer vs half-integer spin) and anisotropy by tuning the species of the magnetic ion and the local crystallographic environment and exchange properties through the nonmagnetic ligand, X~\cite{Coak20:32,Basnet22:4,Olsen21:54, Chittari16:94,Sen20:2,Petkov22:34}.

In this paper, we discuss the nature of the spin fluctuations in the Fe$^{2+}$ compounds, (FePSe$_{3}$ and FePS$_{3}$), with specific emphasis on the former compound in light of recent comprehensive experimental results~\cite{Chen24:9} [we discuss these results in the context of available measurements of FePS$_3$ in Appendix~\ref{appendix:FePS3}]. It has been suggested that the spin excitations in the correlated paramagnetic regime are consistent with transitions between quantum-correlated hexagonal spin clusters, as discussed in the context of large-$S$ pyrochlore Heisenberg antiferromagnets~\cite{Lee02:418,Gao18:97,Tomiyasu08:101,Tomiyasu13:110}. Here, we detail an alternative explanation as classical spin dynamics exhibiting non-linearities from thermal fluctuations. This interpretation overcomes the explanatory difficulty that thermal fluctuations are antithetical to forming spatially extended entangled clusters. Moreover, our finite temperature calculations shed further light on the nature of the order-disorder phase transition in FePSe$_3$.

This paper is organized as follows. We begin with a brief summary of the existing experimental and theoretical results on the FePX$_{3}$ family of compounds. We then examine the low-temperature dynamics in these systems using traditional linear spin wave theory, showing good agreement with the published spectra. We then focus on temperature-dependent fluctuations in this system and demonstrate, using classical dynamics, that the published neutron scattering results are consistent with classical nonlinearities due to the elevated temperature. Finally, we discuss the implication of our results to the understanding of spin-anisotropic two-dimensional magnets.

\section{Exchange Model}
\begin{figure}
    \centering
\includegraphics[width=\linewidth]{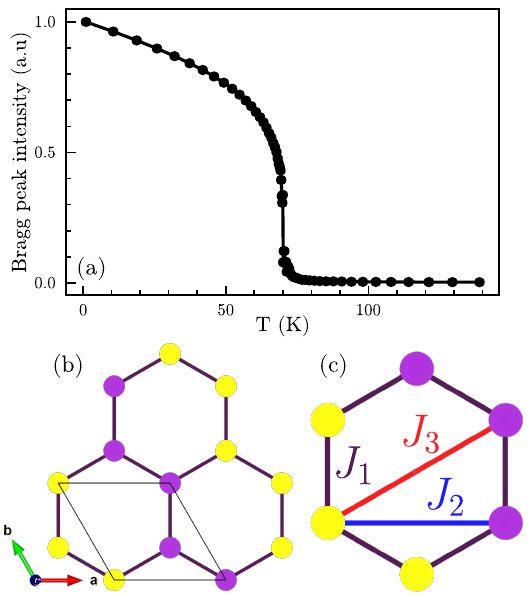}
    \caption{(a) The magnetic order parameter corresponding to $\boldsymbol{k}=(\frac{1}{2},0,\frac{1}{2}$) for the biquadratic (BQ) model of FePSe$_3$ simulated using Langevin dynamics on a (24,24,2) supercell. A sharp phase transition is observed at $T_N^{\mathrm{cl}}=70$ K. (b) The in-plane magnetic structure with up (down) spins in purple (yellow). (c) Definitions of the first three nearest neighbor couplings. }
    \label{fig:order_parameter}
\end{figure}

The crystal structure of FePSe$_{3}$ belongs to the R$\overline{3}$ rhombohedral space group with $C_3$ point group symmetry for Fe$^{2+}$ ions. This contrasts with FePS$_{3}$ which has broken $C_3$ symmetry and adopts the monoclinic $C2/m$ space group. This has ramifications for the allowed exchange models [see Fig.~\ref{fig:order_parameter}(c)], with all nearest neighbor bonds on the honeycomb lattice ($J_1$) being symmetry equivalent.

The measured magnetic moment for FePSe$_3$ is close to the spin-only value, suggesting that the strong trigonal distortion largely quenches the orbital moment; this motivates a traditional treatment of the local spin space as purely dipolar (using coherent states of SU$(2)$) in the absence of off-diagonal magnetic exchange interactions, with $S=2$. Still, small discrepancies between the experimentally measured spectra and predictions from theoretical spin models remain. Recently, in the case of FePS$_3$, it has been suggested that these discrepancies arise due to a partial mixing of excited $J=2$ and $J=3$ spin-orbital levels since an orbital degeneracy exists in the $t_{2g}$ levels of Fe$^{2+}$~\cite{Dhakal24:9,wei24:preprint}. 

We now discuss the exchange model of FePSe$_{3}$ [further comments on FePS$_{3}$ can be found in Appendix~\ref{appendix:FePS3}]. In FePSe$_3$, the magnetic ordering wavevector $\boldsymbol{k}=(\frac{1}{2},0,\frac{1}{2})$ describes a zig-zag order in the plane and antiferromagnetic out of plane order. As shown in Ref.~\onlinecite{Chen24:9}, the out of plane dispersion is significantly weaker than the in-plane dispersion, we therefore set the nearest-neighbor out of plane coupling $J_c$ equal to a small value ($J_c= 0.01$ meV) to achieve the correct ordering wavevector. We consider the two models of Ref. \onlinecite{Chen24:9} consistent with the crystal symmetry to describe the in-plane magnetic interactions. The first, labeled BL (bilinear), is a $J_1$-$J_2$-$J_3$ model with single-ion anisotropy, $D$. The second, labeled BQ (biquadratic), is a $J_1$-$J_2$-$J_3$-$K_1$ model, where $K_1$ is a nearest neighbor ferromagnetic biquadratic interaction. This last term introduces inequivalence in the exchange between different Bravais lattices~\cite{Wildes20:127} and may originate from magnetoelastic coupling or from the presence of a finite orbital moment leading to non-zero multipolar exchange as discussed in Ref.~\onlinecite{Chen23:5} and recently in Ref.\onlinecite{Dhakal24:9} in the context of FePS$_3$. The Hamiltonian can thus be written as 
\begin{equation}
    \hat{\mathcal{H}}=\frac{1}{2}\sum_{i,j }J_{ij}\hat{\mathbf{S}}_i \cdot \hat{\mathbf{S}}_j -D\sum_i (\hat{S}^{z}_i)^{2}+\frac{1}{2}\sum_{ i,j}K_{ij} (\hat{\mathbf{S}}_i \cdot \hat{\mathbf{S}}_j)^{2}
\end{equation}
\noindent with the parameters for models 1-3 given in Table~\ref{table:exchanges}. We use the fitted values from Ref. \onlinecite{Chen24:9} to model FePSe$_3$.

Similar Hamiltonians are reported for the two FePX$_3$ systems. We only present calculations for FePSe$_3$ in this work, but for completeness, we list the proposed parameters for FePS$_3$ Table~\ref{table:exchanges} based on Refs.~\cite{Wildes20:127,Chen24:9}. Whilst the low temperature Hamiltonian parameters are similar for both compounds, subtle differences may arise between the two compounds due to the different atomic numbers of the chalcogen ions, which affects both the strength of spin-orbit coupling and orbital overlap. Furthermore, differences in the magnetoelastic properties may play a role close to the ordering transition which is accompanied by an observable structural transition in FePS$_3$~\cite{Jernberg84:46}.

Throughout this paper, we utilize the Renormalized Classical Spin (RCS) theory~\cite{Dahlbom23:preprint} to describe the spin dynamics. This approach more accurately describes single-ion anisotropies and biquadratic exchange than the traditional large-$S$ theory, because it directly uses the classical limit of higher-order multipolar spin operators, rather than taking the classical limit of each individual spin operator. For quadratic anisotropies, the correspondence between the two methods is exact in the limit $S\to \infty$, but a correction factor must be applied for finite $S$. In the case of biquadratic exchange, the RCS theory generates a correction that is \textit{bilinear} in spin operators and must be accounted for in the strength of exchange interactions to match the fitted spectrum in Ref~\onlinecite{Chen24:9}. This term is independent of $S$ and so even for small values of $1/S$, this correction to the large-$S$ limit remains. The correspondence between large-$S$ and RCS is summarized in Appendix~\ref{appendix:RCS}. These renormalization factors are distinct from the quantum corrections generated by a mean-field averaging of higher-order $1/S$ corrections to the Hamiltonian, which have been neglected from our analysis. The origin of the latter corrections are a result of decoupling of terms of magnon scattering terms~\cite{Zhitomirsky13:85}.

\section{Static Correlations}
We start by determining the ground state spin configuration of FePSe$_3$ by using the (adequately classically renormalized) exchanges listed in Table~\ref{table:exchanges}. We do so in an unbaised manner by performing Langevin dynamics simulations using the \textsc{Sunny.jl} package~\cite{Dahlbom25:preprint,Dahlbom22:106,Dahlbom22:106_2,Zhang21:104} [See Appendix~\ref{appendix:LLD} for details]. Starting from the high temperature (random) limit, we thermalize a $24 \times 24 \times 2$ supercell of spins (in the hexagonal setting) followed by gradient descent to find the $T=0$ magnetic structure. We verify that for both FePSe$_3$ models (BL and BQ), the ground state magnetic order is the $\boldsymbol{k} = (\frac{1}{2},0,\frac{1}{2})$ zig-zag structure, or its domains generated under $120^{\circ}$ rotations. In the case of FePS$_3$, we find the $\boldsymbol{k} = (0,1,\frac{1}{2})$ structure, which is the same zig-zag arrangement as observed in FePSe$_3$ up to a redefinition of the basis vectors as appropriate for the $C2/m$ monoclinic space group. 

To further explore the nature of the magnetic ordering transition in FePSe$_3$ we calculate the magnetic order parameter as a function of temperature using the same Langevin dynamics approach. Care was taken to ensure samples were taken after sufficient time to thermalize and decorrelate the spin configurations [Appendix ~\ref{appendix:decorrelation}]. We observe a sharp transition at $T_N$ followed by a gradual saturation as $T\to 0$ [Appendix~\ref{appendix:TN}]. 

A well-known deficiency of classical simulations is the failure to capture the first-order nature of the transition. This fact was pointed out in Ref.~\onlinecite{Dahlbom24:109} and can be overcome by restricting the dimensionality of the fluctuations manifold by considering Ising spins (and thus taking $J_{i}^{\alpha\beta} \to 0 $ for $\alpha,\beta \neq z$). This leads to a better accounting of the first-order nature of this transition.  However, such a model does not accurately describe the dynamics in systems with manifestly Heisenberg spins and exchange. Based on the fitting to the low temperature neutron data alone, the presence of Ising-like exchange anisotropy  cannot be ruled out entirely, and the inclusion of such anisotropies may lead to a better description of the ordering transition.

The overestimation of fluctuations close to the ordering temperature also manifests in an underestimation of $T_{\rm N}$. In our classical simulations, we find that $T^{\mathrm{cl}}_N = 70$ K, whilst experimentally, the ordering temperature was determined to be $T^{\mathrm{exp}}_{\rm N} = 110$ K. This underestimation can be understood in the context of Weiss mean field theory which predicts $T_N$ to differ by a factor of $(S+1)/S$ between the quantum and classical models. Notwithstanding these limitations, our classical simulations correctly capture the sharp order parameter increase between a paramagnetic state and a zig-zag antiferromagnetic state below $T_{\rm N}$. Another contributing factor to these discrepancies is the use of Gaussian white noise which does not account for the necessary crossover to quantum (Bose-Einstein) statistics to describe spin fluctuations of the model at low temperatures. The implementation of colored noise~\cite{Savin12:17} has been shown to accurately capture the low temperature thermodynamic properties of magnetic systems~\cite{Barker19:100}, but, as far as we know, a practical implementation of such thermostat in classical dynamics simulations is still lacking. Nevertheless, traditional Landau-Lifshitz dynamics has been shown to qualitatively reproduce the experimentally measured finite temperature dynamics of interacting spin systems with an appropriate rescaling of the temperature [see Refs.~\cite{Do23:8,Dahlbom24:109,Park24:6} for example].

\section{Linear Spin Wave Spectrum}
\begin{figure}[b]
    \centering
    \includegraphics[width=\linewidth]{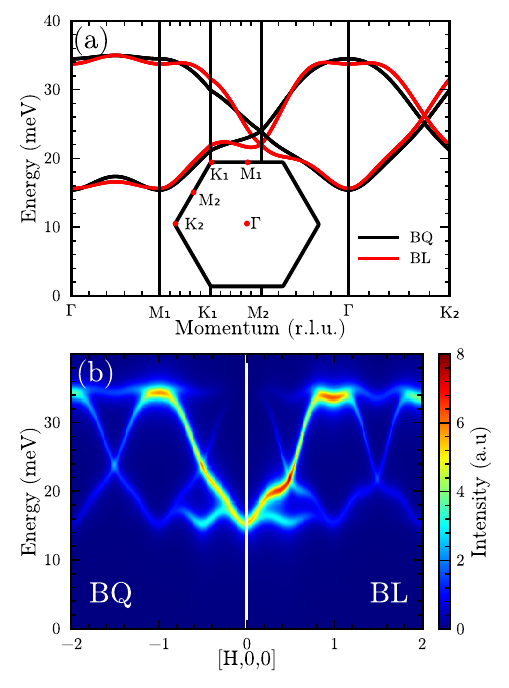}
    \caption{Zero-temperature dynamical response of FePSe$_3$ predicted by linear spin-wave theory through high symmetry points of the Brillouin Zone, for both the bilinear (BL) and biquadratic (BQ) exchange models using the RCS. High symmetry points in the Brillouin zone are labeled in the inset. (a) Dispersion curves (b) Neutron scattering intensity averaged over all three equivalent domains.}
\label{fig:LSWT_FePSe3}
\end{figure}

We now turn to the $T=0$ magnetic excitations expected in the ordered phase of FePSe$_3$. In Fig.~\ref{fig:LSWT_FePSe3}(a), we employ linear spin-wave theory to plot the magnon dispersion relation through a path connecting high symmetry points of the hexagonal Brillouin zone, for the two FePSe$_3$ models considered in Ref~\onlinecite{Chen24:9}. Small differences are observed between the BL and the BQ models, which are most apparent near the band crossing at the M$_2$ point of the Brillouin zone. In Fig.~\ref{fig:LSWT_FePSe3}(b), we show the corresponding neutron scattering intensity as a function of momentum transfer ${\bf Q}$ and energy transfer $E$, which reveals additional small differences in intensity between the two models, the experimental discrimination of which is likely challenging. 

A complete understanding of the spin dynamics of FePSe$_3$ requires modeling dispersive higher energy spin-orbital excitations observed around $\sim$ 39 and 44 meV. In principle, this can be achieved by representing the local Hilbert space using coherent states of SU$(5)$ and including bilinear exchange interactions that directly couple components of the resulting magnetic multipoles. However, the coupling between high- and low-energy excitations is small, leading to a negligible neutron scattering intensity for these modes. Including mixing between higher spin-orbital states could potentially give rise to non-zero matrix elements between the ground-state and higher energy excited states; neutron scattering measurements at higher energy transfer may shed light on this question. Given the number of possible exchange interactions between SU(5) coherent states, we instead rely on a purely dipolar model for the spins in FePSe$_3$.

\begin{table}
\caption{ Parameters used in our calculations, adapted using the RCS to be consistent with \textcite{Chen24:9}. A small interchain coupling has been added to reproduce the reported magnetic ground state. $^{**}$Reported value includes the RCS renormalization. }
\begin{ruledtabular}
\begin{tabular}{c|cc}
Bond & BL & BQ \\
\hline\hline 
$J_1$ & -2.30~  &-1.32~~ \\
$J_2$ & -0.23~  &~0.12~~ \\
$J_3$ & ~2.01~  &~1.28~~ \\
$J_c$ & ~0.01$^{*}$  &~0.01$^{*}$~ \\
$K_1$ & ~0.00~      &-0.22$^{**}$ \\
$K_2$ & ~0.00~     &~0.00~~ \\
$D$   & 3.65~  &~3.08~~ 
\end{tabular}
\end{ruledtabular}
\label{table:exchanges}
\end{table}
A cruder but more practically feasible approach is to introduce biquadratic exchange between purely dipolar spins, as it has been suggested to provide a better description of experimental data in FePSe$_3$~\cite{Chen24:9}, and similar arguments have been made for FePS$_3$~\cite{Wildes20:127}. We therefore restrict our further discussion to the BQ model. The inclusion of biquadratic coupling in $2D$ magnetism has been discussed at length in the literature (see, for example, Refs.~\cite{Turner09:80,Slonczewski91:67,Fedorova15:91,Gutzeit22:13,Kreisel20:12}). Presence of this term does not protect order in $2D$~\cite{Krzeminski76:74} but is often required for a strong agreement with experiment, particularly in the extreme two-dimensional limit~\cite{Olsen24:11,Kartsev20:6}.

\section{Temperature-dependent spin fluctuations}
To model the spin dynamics of FePSe$_3$ at finite temperature, including in the paramagnetic regime above $T_{\rm N}$, we turn to classical (Landau-Lifshitz) dynamics simulations for dipoles using the renormalized parameters of Tab.~\ref{table:exchanges}. A known challenge in such simulations is the correct treatment of the classical-to-quantum correspondence to faithfully describe the dynamics over a broad range of temperatures~\cite{Dahlbom23:preprint}. One approach relies on enforcing the zeroth moment sum-rule of the dynamical structure factor, $\sum_\alpha\iint d\omega d\mathbf{q} S^{\alpha\alpha}(\mathbf{q},\omega) =S(S+1)$, which we can enforce by requiring that $\sum_{i}\langle \hat{\mathbf{S}}_{i}^{2} \rangle=N_{s} S(S+1)$ is satisfied by our classical dipoles at each simulation temperature. While the correspondence principle for the harmonic oscillator ensures the correct sum rule in the low temperature (harmonic) limit, a temperature-dependent rescaling factor for the dipole moments is, in principle, required for finite temperatures, reaching $\sqrt{1+1/S}$ as $T\to \infty$~\cite{Dahlbom23:preprint}. For $S=2$, $\kappa(T\to \infty)\approx 1.22$~\cite{Zhang21:104,Dahlbom23:preprint}. Since this renormalization factor is modest, we neglect this from our analysis.

\begin{figure*}
    \centering
    \includegraphics[width=\linewidth]{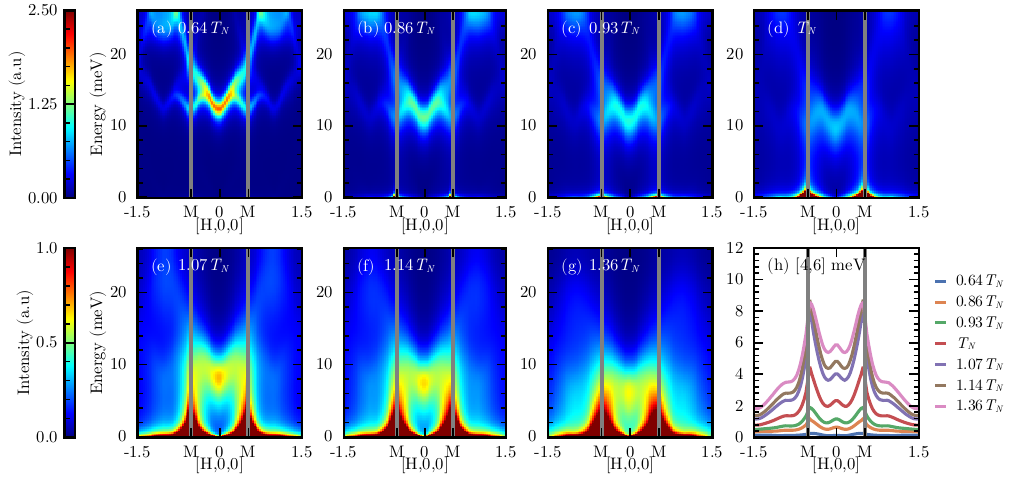}
    \caption{(a-g) Simulated inelastic neutron scattering intensity represented as $(\mathbf{Q},E)$ slices taken at increasing temperatures. Slices in the second row are plotted on a smaller intensity scale to aid visibility. Gray vertical lines indicate the location of the M points in the first Brillouin zone. Numerical divergences that appear as artifacts in the ordered phase due to the finite length of the real-time trajectories are deliberately obscured by these gray lines. (h) Constant energy cuts at $E=5$ meV taken through the $(\mathbf{Q},E)$ slice plotted in (a-g).}
    \label{fig:QEslices}
\end{figure*}

Simulation results are plotted in Fig.~\ref{fig:QEslices}. The calculations are once again domain-averaged, and are integrated over $L=[-2,2]$ out of plane. At $T=0.64 T_{N}^{\mathrm{cl}}$, a dispersive mode is still visible, albeit with a broadened lineshape compared to the zero-temperature limit. The broad linewidth indicates a finite lifetime induced by thermal nonlinearities of the classical dynamics. At $T=0.86 T_{N}^{\mathrm{cl}}$ the dispersive modes are still present with a reduction in intensity reflecting a redistribution of spectral weight to diffuse quasi-elastic features. A further reduction of the anisotropy gap is caused by a thermal fluctuation induced reduction of the ordered moment. As $T$ crosses the ordering temperature [Fig.~\ref{fig:QEslices}(d)], broad features extend from the wavevectors associated with the antiferromagnetic ordering wavevector, indicating the presence of short-range antiferromagnetic correlations. These extend up and join the spin wave branch at $T_N$. The broad columnar features persist to higher temperatures [Figs.~\ref{fig:QEslices}d-f] with the increased temperature redistributing intensity to the columns protruding from the locations of the magnetic Bragg peaks. The scattering remains structured at temperatures far exceeding the magnetic ordering temperature [Fig.\ref{fig:QEslices}(f)], reproducing the signal observed experimentally~\cite{Chen24:9}. To further clarify this temperature evolution, we calculate a constant energy cut at $E\!=\!5$~meV. At $T=0.64 T_{N}^{\mathrm{cl}}$ no signal is observed in this range. As temperature is increased intensity accumulates at the antiferromagnetic wavevector leading to broad features across the Brilloin zone at $T=1.07 T_{N}^{\mathrm{cl}}$, $T=1.08 T_{N}^{\mathrm{cl}}$ and $T=1.14 T_{N}^{\mathrm{cl}}$. 

Present in our calculation is a region of high intensity at the zone center above $T_N$ which can be attributed to incipient short range spin wave excitations which develop close to $T_N$ and form the bottom of the magnon band seen in the ordered phase, where the density of thermally excited magnons is greatest [Fig.~\ref{fig:QEslices}(a)]. In our calculation this feature is smeared out at higher temperatures. The evidence for this excess of intensity at the zone center in the experimental data~\cite{Chen20:101} is inconclusive.  

\begin{figure}
    \centering
    \includegraphics[trim={5.5cm 0 5.5cm, 0},clip,width= \linewidth]{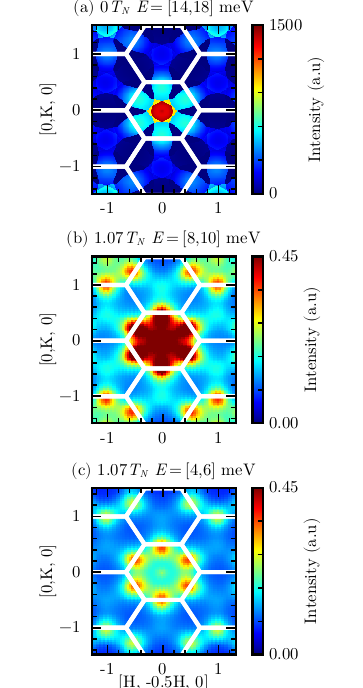}
    \caption{(a) Low-temperature constant energy slice at $E=16$ meV calculated using linear spin-wave theory. Constant energy slices through the ($\mathbf{Q},E$) simulation results of Fig.~\ref{fig:QEslices} at $1.07T_N$ with energy transfers of (b) $E=9$~meV and (c) $E=5$~meV calculated using Landau-Lifshitz dynamics. }
    \label{fig:constE}
\end{figure}

We now consider the in-plane magnetic fluctuations by plotting the simulated inelastic neutron scattering intensity at constant energy. We begin in the ordered phase at base temperature. Using linear spin wave theory, we examine a constant energy slice at the bottom of the band [Fig.~\ref{fig:constE}(a)]. Our calculations reproduce the results of Ref.~\onlinecite{Chen24:9} showing the $C_3$ symmetric pattern associated with the excitations within the domain-averaged system.

We then take constant energy slices through the inelastic simulations presented in Fig.~\ref{fig:QEslices} comparable to those presented in Fig. 4 of Ref.~\onlinecite{Chen24:9}. Fig.~\ref{fig:constE}(b) shows a slice near the top of the column of inelastic intensity in Fig.~\ref{fig:QEslices}(d). We observe intensity throughout the Brillouin zone, with bulges at the M points, which are associated with the low-temperature zig-zag antiferromagnet order and the greatest intensity at the $\Gamma$ point, as observed in experiment~\cite{Chen24:9}. Fig.~\ref{fig:constE} shows a constant energy slice taken through the columns protruding from the antiferromagnetic ordering wavevectors [Fig.~\ref{fig:QEslices}]. The accumulation of intensity at the M points, as observed in the data, is captured in the simulation, as is the absence of intensity towards the $\Gamma$ point. 
 
\section{Discussion}

Our simulations demonstrate the remarkable success of semiclassical methods in describing a broad range of excitations beyond harmonic spin waves. In the zero-temperature limit of FePSe$_3$, linear spin wave theory captures the harmonic fluctuations of the dipolar spin moments about the long-range ordered $\boldsymbol{k}=(\frac{1}{2},0,\frac{1}{2})$ magnetic structure. As temperature increases, linear spin wave theory fails to capture classical nonlinearities and the suppression of the ordered moment. To incorporate these effects, we employed Langevin dynamics simulations to describe the stochastic nature of the finite temperature dynamics. Langevin dynamics capture the energy and momentum dependence of the measured spin fluctuations.  

It has previously been suggested that the finite-temperature fluctuations are associated with hexagonal cluster excitations, which drive the magnetic ordering transition by condensing to form the zig-zag antiferromagnetic ground state~\cite{Chen24:9}. The proposed cluster excitations are resonance-like, associated with a local configurational energy above the disordered ground state. One, therefore, expects a decoupling between energy and momentum, with the latter arising due to the spatial extent of the cluster unit. This mechanism has been invoked to explain short-ranged correlations in a variety of systems including Na$_2$Co$_2$TeO$_6$~\cite{Yao22:129} and Ni$_2$Mo$_3$O$_8$~\cite{Gao23:14}. 

One prominent family of systems to which the local cluster model has been applied is the Cr$^{3+}$ spinels, in which $S=3/2$ ions reside on a frustrated pyrochlore lattice. Above a spin-Peierls transition, rings of scattering intensity are observed in momentum space, which have been interpreted as hexagonal loop excitations in real space. Constant energy slices at higher energies display qualitatively different momentum dependence ascribed to different modes of tetramer excitations~\cite{Gao18:97,Tomiyasu13:110}. In MgCr$_2$O$_4$, excitations below the magnetic ordering temperature show a flat mode at low energy, which has been suggested as evidence of the persistence of cluster fluctuations down to low temperature. Whilst these models successfully describe the momentum dependence at a given energy, they do not describe the energy-resolved spectra. In Ref. ~\onlinecite{Bai19:122}, some of us, along with coauthors, demonstrated that above $T_N$, the spectra could be understood semiclassically in terms of harmonic spin wave excitations above the disordered ground state associated with the macroscopically-degenerate classical spin liquid phase. Further to this, in Ref.~\onlinecite{Nassar24:109}, the low energy resonance mode was shown to be dispersive, consistent with coherent spin wave excitations.

Thus, the mechanism underlying the broadened response in FePSe$_3$ differs from the case of a highly frustrated magnet such as MgCr$_2$O$_4$. In FePSe$_3$, the broadening originates from classical nonlinearities of the spin dynamics induced by thermal fluctuations. In MgCr$_2$O$_4$ the broadened response measured just above $T_N$ is primarily a result of harmonic fluctuations averaged over a quasi-degenerate manifold of spin configurations. This difference reflects the distinct origins of the spin disorder in the two systems. In MgCr$_2$O$_4$, $T_N$ is small compared to the exchange interaction scale ($\Theta_W /T_N \approx 30$)~\cite{Watanabe12:86} and hence the nonlinear terms in the equation of motion are small at or just above $T_N$. In contrast, $T_N$ is roughly on the same order of the Weiss constant~\cite{Joy92:46} for FePX$_3$. It is, therefore, to be expected that the nonlinear fluctuations close to $T_N$ are sizeable and play a key role in the dynamics of the latter compound. 

The results presented in this paper suggest that the finite temperature dynamics in FePSe$_3$ can be well described using semiclassical methods. In this picture, the order-disorder phase transition at $T_N$ can be viewed as the overcoming of the anisotropy gap, which protects the long-range magnetic ordering in two dimensions. The sharp nature of the transition makes the experimental observation of the gap closure challenging since the gap rapidly closes as thermal fluctuations overcome the anisotropy energy. Similarities can be observed between the order-disorder transition in FePSe$_3$ and CrI$_3$, a weakly anisotropic ferromagnet for which the phase transition is weakly first order and the gap goes to zero at the transition whilst the spin stiffness remains finite~\cite{Chen20:101}. 

The finite spin stiffness at the phase transition also accounts for the highly structured magnetic fluctuations observed above $T_N$, which arise from strong short-range correlations. The nature of the gap destruction in FePSe$_3$ is somewhat different from the traditional picture of gap closure such as the field induced transition of a quantum antiferromagnet in field, where the singlet condensate is replaced by a magnon condensate as the magnon gap closes. Here, the magnon density is the order parameter, growing continuously as the field increases. In FePSe$_{3}$, thermal fluctuations protruding from the antiferromagnetic wavevector overcome the anisotropy energy abruptly as the temperature increases above $T_N$. In this sense, we can view this as a closure of the anisotropy gap \textit{from below}. 

We note that the first-order nature of the transition, along with the observation of a structural distortion in FePS$_3$~\cite{Jernberg84:46} are suggestive of a magnetoelastic origin for the phase transition in FePS$_3$. To our knowledge, no evidence of such distortion has been observed in FePSe$_3$~\cite{Wiedenmann81:40}, and the success of our model in describing the fluctuations at, and above, $T_N$ suggests that magnetoelasticity may not drive the order-disorder transition. Measuring the magnetic fluctuations close to the ordering transition in FePS$_3$ would be interesting to compare with  FePSe$_3$. The situation in FePSe$_3$ appears reminiscent of FeI$_2$ which exhibits a first order thermodynamic phase transition without the observation of a structural transition~\cite{Petitgrand80:41}. There are further similarities between the dynamics in FePSe$_3$ and FeI$_2$, where the spectrum remains gapped as $T_N$ is approached~\cite{Dahlbom24:109}.

\section{Conclusion}
In this paper, we have presented semiclassical calculations describing the magnetic order and dynamics of the 2D van der Waals antiferromagnetic systems FePSe$_3$. Using a combination of linear spin wave theory and Landau-Lifshitz dynamics for dipolar spins, we have reproduced the zero-temperature spin wave excitations and the short-range fluctuations at finite temperature with a common model. Our simulations point to the overcoming of the anisotropy energy by thermally-induced antiferromagnetic fluctuations as the origin of the first-order magnetic order-disorder phase transition. We speculate that these results are general to a broad range of large-$S$ 2D van der Waals magnets for which a strong easy anisotropy sets the magnetic ordering temperature. These results further motivate careful temperature-dependent spectroscopic and diffraction measurements in FePSe$_3$.

\begin{acknowledgments}
    H.L was funded by a Research Fellowship from the Royal Commission for the Exhibition of 1851. M.M acknowledges support from U.S. Department of Energy, Office of Science, Basic Energy Sciences, Materials Sciences and Engineering Division under award DE-SC-0018660. H.L acknowledges useful discussions with L. Chen, J.-H. Chung, R. J. Birgeneau, P. Dai, A. R. Wildes, C. D. Batista, D. Dahlbom, K. Barros and X. Bai. The research data supporting this publication can be accessed at http://dx.doi.org/10.6084/m9.figshare.29077010~\cite{opendata}.
\end{acknowledgments}
\appendix
\section{FePS$_3$}
\label{appendix:FePS3}
\begin{table}
\caption{Exchange parameters (in meV) adapted from \textcite{Wildes20:127} using the RCS to be consistent with conventions used in calculations for FePSe$_3$ in this paper. $^{*}$A small interchain coupling has been added to reproduce the reported magnetic ground state. Exchange couplings from include a factor of two to account for the difference in Hamiltonian definition.}
\begin{ruledtabular}
\begin{tabular}{l| cp{2in}}
$J_1$ & -2.02  \\
$J_2$&  ~0.30 \\
$J_3$ &  ~0.9\\
$J_c$ & ~0.01$^{*}$ \\
$K_1$ &-0.20 \\
$K_2$ & -0.04\\
$D$   &  ~3.36 
\end{tabular}
\end{ruledtabular}
\label{table:FePS3}
\end{table}
The crystal structures of FePS$_{3}$ differs from FePSe$_3$ having a broken $C_3$ symmetry and crystallizing in the monoclinic $C2/m$ space group. The broken $C_3$ symmetry splits the nearest-neighbor bonds on the honeycomb lattice into inequivalent zig-zag, $J_{1a}$ and leg, $J_{1b}$ bonds. The subtle nature of the distortion has nonetheless motivated models where both $J_{1a} = J_{1b}$ and $J_{1a} \neq J_{1b}$ in FePS$_{3}$~\cite{Lancon16:94, Wildes20:127}, however, the most successful fits have employed models where $J_{1a} = J_{1b}$.

A $J_1$-$J_2$-$J_3$-$K_1$-$K_2$ model with single-ion anisotropy best describes the inelastic neutron scattering data~\cite{Wildes20:127}, with parameters qualitatively similar to the fitted values for FePSe$_3$~\cite{Chen24:9} including with a sizable ferromagnetic nearest-neighbor and antiferromagnetic next-nearest-neighbor Heisenberg exchange, a strong uniaxial anisotropy, and a ferromagnetic biquadratic nearest-neighbor exchange. For completeness, we quote the fitted parameters for FePS3 in Table~\ref{table:FePS3}.
\section{Renormalized Classical Spin Theory}
\label{appendix:RCS}
In order to compare the calculations in the literature with our work, we rescale the Hamiltonian parameters to account for the corrections to leading order in $1/S$ according to Ref.~\onlinecite{Dahlbom23:preprint}. The correction to the single ion anisotropy is 
\begin{subequations}
\begin{gather}
    D\to \frac{D}{r}\\
    r = 1-\frac{1}{2S}.
\end{gather}
\end{subequations}
\noindent The biquadratic exchange acquires both a renormalization and a bilinear correction. To recover the result in the literature we add back the bilinear correction and renormalize the strength of the biquadratic exchange
\begin{subequations}
    \begin{gather}
        K_{ij}(\mathbf{\hat{S}}_{i}\cdot\mathbf{\hat{S}}_{j})^{2}\to \frac{K_{ij}}{r}(\mathbf{\hat{S}}_{i}\cdot\mathbf{\hat{S}}_{j})^{2}+\frac{1}{2}K_{ij}\mathbf{\hat{S}}_{i}\cdot\mathbf{\hat{S}}_{j}\\
        r = 1-\frac{1}{S}+\frac{1}{4S^2}.
    \end{gather}
\end{subequations}\\

\section{Landau Lifshitz Dynamics}
\label{appendix:LLD}
The finite temperature dynamics are simulated the Landau Lifshitz equations with noise and damping terms that are chosen to enforce the fluctuation-dissipation theorem. In this appendix, we reproduce a number of the key equations for completeness.

The Landau Lifshitz equation, for the three component spin dipole, $\mathbf{S}_j$ with damping and noise is given by
\begin{equation}
\label{eq:LLD}
    \frac{d\mathbf{S}_j}{dt}=-\mathbf{S}_j \times\left(\boldsymbol{\xi}_j+ \frac{\partial \mathcal{H}}{\partial \mathbf{S}_j}-\lambda \mathbf{S}_j \times \frac{\partial \mathcal{H}}{\partial \mathbf{S}_j}\right).
\end{equation}
\noindent The noise term is chosen to be a Gaussian white noise with
\begin{gather}
    \langle \xi_j^\alpha(t)\rangle = 0\\
     \langle\xi_i^\alpha(t) \xi_j^\beta(t')\rangle = 2\lambda k_B T\delta_{ij}\delta_{\alpha\beta}\delta(t-t').
\end{gather}
\noindent Equation~\ref{eq:LLD} was numerically integrated using the explicit Heun method in the \textsc{Sunny.jl} spin dynamics package~\cite{Dahlbom25:preprint}. Interested readers are referred to Refs.~\cite{Dahlbom22:106,Dahlbom22:106_2,Dahlbom25:preprint} for further details.
\section{Decorrelation}
\label{appendix:decorrelation}
When performing Langevin dynamics simulations, measurements must be taken once the system is sufficiently decorrelated. In order to estimate the decorrelation time, the autocorrelation function
\begin{equation}
    \mathcal{A}(t)=\frac{\langle\hat{\mathcal{O}}_i\hat{\mathcal{O}}_{i+t}\rangle-\langle \hat{\mathcal{O}}_i\rangle^{2}}{\langle\hat{\mathcal{O}}_i^{2}\rangle-\langle \hat{\mathcal{O}}_i\rangle^2}
\end{equation}
\noindent was calculated, and the integrated decorrelation time \begin{equation}
    \tau_{\mathrm{int}}=\frac{1}{2}+\sum_t \mathcal{A}(t)
\end{equation}
\noindent was determined for a series of temperatures between base temperature and 120 K. It was determined that $\tau_{\mathrm{int}}<530$ Langevin steps for all temperatures and so the time between measurements was set to 2$\tau_\mathrm{int}=1060$ to ensure sufficiently decorrelated samples at all temperatures.\\

\section{Determining $T_{N}^{\mathrm{cl}}$}
\label{appendix:TN}
We calculate the classical ordering temperature, $T_{N}^{\mathrm{cl}}$ for each model under consideration by integrating about the Bragg peak as a function of temperature and numerically differentiating the order parameter as a function of temperature. The site of the discontinuity was taken to be the classical ordering temperature $T_{N}^{\mathrm{cl}}$. Calculations were performed on an $24\times24\times2$ supercell with 1000 samples at each of 59 temperatures, spaced exponentially from both directions to provide dense coverage around the transition temperature.

\bibliography{references.bib}

\begin{thebibliography}{74}
\bibitem [{ope()}]{opendata}%
  \BibitemOpen
  \href@noop {} {\bibinfo {title} {Dataset to be added.}}\BibitemShut {Stop}%
\end{thebibliography}
\end{document}